\documentclass[oneside]{article}
\usepackage[T1]{fontenc}
\usepackage{authblk}
\usepackage{float}
\usepackage{amsmath}
\usepackage{graphicx}
\usepackage{xfrac}
\usepackage[english]{babel}
\usepackage{lmodern}
\usepackage[T1]{fontenc}
\usepackage[latin9]{inputenc}
\usepackage{mathtools}
\usepackage{graphicx}
\usepackage{esint}
\usepackage[unicode=true,
 bookmarks=false,
breaklinks=false,pdfborder={0 0 1},backref=false,colorlinks=false]
 {hyperref}
\usepackage[a4paper, total={210mm,297mm},margin=2.0cm]{geometry}

\usepackage{datetime}
\newdate{date}{12}{5}{2016}

\title{Chimera Multiscale Simulation of Complex Flowing Matter}

\author[1,2]{Sauro Succi \thanks{Electronic address: \texttt{s.succi@iac.cnr.it}; Corresponding author}}

\affil[1]{Istituto per le Applicazioni del Calcolo CNR, Via dei Taurini 19, 00185 Rome, Italy}
\affil[2]{Harvard Institute for Applied Computational Science, Cambridge, Massachusetts, USA}

\date{\displaydate{date}}

\begin{document}

\maketitle

\begin{abstract}
We discuss a unified mesoscale framework for the simulation 
of complex states of flowing matter across scales of motion which requires 
no explicit coupling between different macro-meso-micro levels. 
The idea is illustrated through selected examples of 
complex flows at the micro and nanoscale. 
\end{abstract}


\section{Multiscale without hand-shaking: the chimera concept}

Multiscale modeling is the natural response to the hierarchical organization 
of the  interactions which shape up the complexity of the world around us.
Traditional multiscale computing is based on the notion of {\it overlapping} between the
four basic levels of the BBGKY hierarchy, macro, meso, micro and quantum \cite{WEINAN}.
The overlap between these levels is sustained by the continued advances
in computer hardware and modelling techniques.
Yet, the gaps remain daunting, easily ten decades in space (say Angstroms to meters) and
twice as many in time (femtoseconds to hours). In addition, the practical implementation
of the hand-shaking interfaces across the different levels remain laborious and technically 
intensive \cite{MAAD}, \cite{CISE}, \cite{FILIPPA}.
Under such state of affairs, it is highly desirable to explore different
multiscale strategies based on a {\it unified} treatment of multiple levels of the
hierarchy without handshaking interfaces.
In this context, the concept of overlapping is replaced by the idea of {\it morphing}, i.e. the
unified scheme should be able to morph into the desired level, say micro and continuum, wherever and
whenever needed "on demand".
This is what we shall refer to as to {\it chimera} multiscale.
At variance with the original mythological meaning, to us the chimera is not a 
monster, but a graceful and flexible creature! 


With the above premises, it is clear that mesoscale represenations, and particularly kinetic theory, lie 
at a vantage point to implement the chimera strategy, since they offers
 a natural intermediate description between the micro and macro descriptions. 
Indeed, Boltzmann kinetic theory is based on a continuum field,
the probability distribution function, propagating along particle-like trajectories.
This offers a natural particle-field duality which can prove pretty useful for multiscale purposes.
For instance, instead of coupling continuum hydrodynamics to atomistic models, the mesoscale 
chimera would morph, so to say, into both these levels by suitable "mutations" 
which take it close to Navier-Stokes on the upper end and to Newtonian mechanics on the lower one.
By doing so, the need for handshaking between different levels is lifted altogether.
Of course there are limits to this strategy, 

In the sequel of this paper, we shall provide a cursory description
of the mutations which promote this chimera behaviour, and discuss its potential
and limitations.

\section{Universality and Molecular Individualism: Weakly Broken Universality}

Before doing so, let us address a natural question:
how far can we push the (unconventional) idea of chimera multiscale? 
In other words, how close can one take the mesoscale
description to the upper macro and lower micro levels, before hitting the ceiling (floor)
where handshaking can no longer be postponed? 
The ceiling part is solid: it is known that, under the appropriate conditions 
(weak departure from local equilibrium), the Boltzmann equation converges to the
Navier-Stokes equation. Here the morphing is guaranteed to be convergent.
The floor side, on the other hand, tells another story.
It is indeed obvious that Boltzmann's kinetic theory cannot be 
made equal to molecular dynamics. 
How close the two can be brought together is controlled by the degree of 
universality exposed by each given problem.
More precisely, by the degree of {\it broken} universality, meaning by this the departure
from universal behaviour, the one associated with the continuum description of moving
matter. On the other hand, complex moving matter is characterized, and we might say even 
{\it defined}, by the coexistence of Universality and Molecular Individualism, to an extent which
changes from problem to problem.
It is hereby surmised that the chimera approach is well suited to capture the fascinating
physics stemming from such coexistence.
If universality is only weakly broken (WBU),  i.e. the physics under exploration
requires more descriptors, fields and parameters, than continuum hydrodynamics, but 
still no strict knowledge of molecular details.
Under such conditions,  the Chimera approach
can provide major computational returns.
Otherwise, conventional hand-shaking cannot be helped.
In the following, we shall substantiate the above program by means of concrete examples
based on a specific mesoscale technique, known as Lattice Boltzmann (LB) method, which we
next proceed to briefly illustrate.

\section{Lattice Boltzmann in a nutshell}

The LBE, in single-relaxation form for simplicity \cite{BGK}, reads as follows \cite{LBE1,LBE2,LBE3,LB2038}
\begin{equation}
\label{LB}
f_i(\vec{x}+\vec{c}_i \Delta t, \vec{c}_i;t+ \Delta t) - f_i(\vec{x},\vec{c}_i;t) =
\omega(f_i^{eq}-f_i) + \gamma S_i
\end{equation}
where the index $i$ labels the set of discrete velocities.
In the above, $f_i^{eq}$ denotes a lattice version of the Maxwell-Boltzmann local
equilibrium, parametrised by the fluid density $\rho=\sum_i f_i$ and velocity
$\vec{u} = \sum_i \vec{c}_i f_i /\rho$, both functions of space and time.

The parameter $\omega = \Delta t/\tau_c$ is the ratio of
free-streaming version collision timescales, $S_i$ is
a source term associated with external/internal forces
and $\gamma = \Delta t/\tau_S$ is a measure of the strength
of the external versus collisional relaxation.

In other terms, $\omega$ measures the strength of collisions versus free
propagation, so that $\omega \to \infty$ connotates the strongly-coupled
hydrodynamic regime, whereas $\omega \to 0$ denotes the weakly-coupled
non-hydrodynamic regime leading to free molecular streaming.
Likewise, $\gamma \to \infty$ denotes strong interaction with external
(or internal) driving sources, while $\gamma \to 0$ indicates weak
coupling to the source.
For the case of sources of momentum, the parameter $\gamma$ is 
proportional to the Froude number
$Fr \equiv \frac{v_{th} \tau_c}{a}$,
where $v_{th}$ is the thermal speed and $a=F/m$ the acceleration
due to the force $F$.
Linear stability imposes the constraint
$0< \omega < 2$,  which secures positive kinematic viscosity through the relation
$
\nu = c_s^2 (1/\omega -1/2) \frac{\Delta x^2}{\Delta t}.
$
Likewise, the parameter $\gamma$ should be taken well below the unit
value, on pain of ruining the stability of the scheme, as we are going to 
detail in the following.

\begin{figure}[h!]
\label{D2q9D3q19}
\centering
\includegraphics[width=24pc]{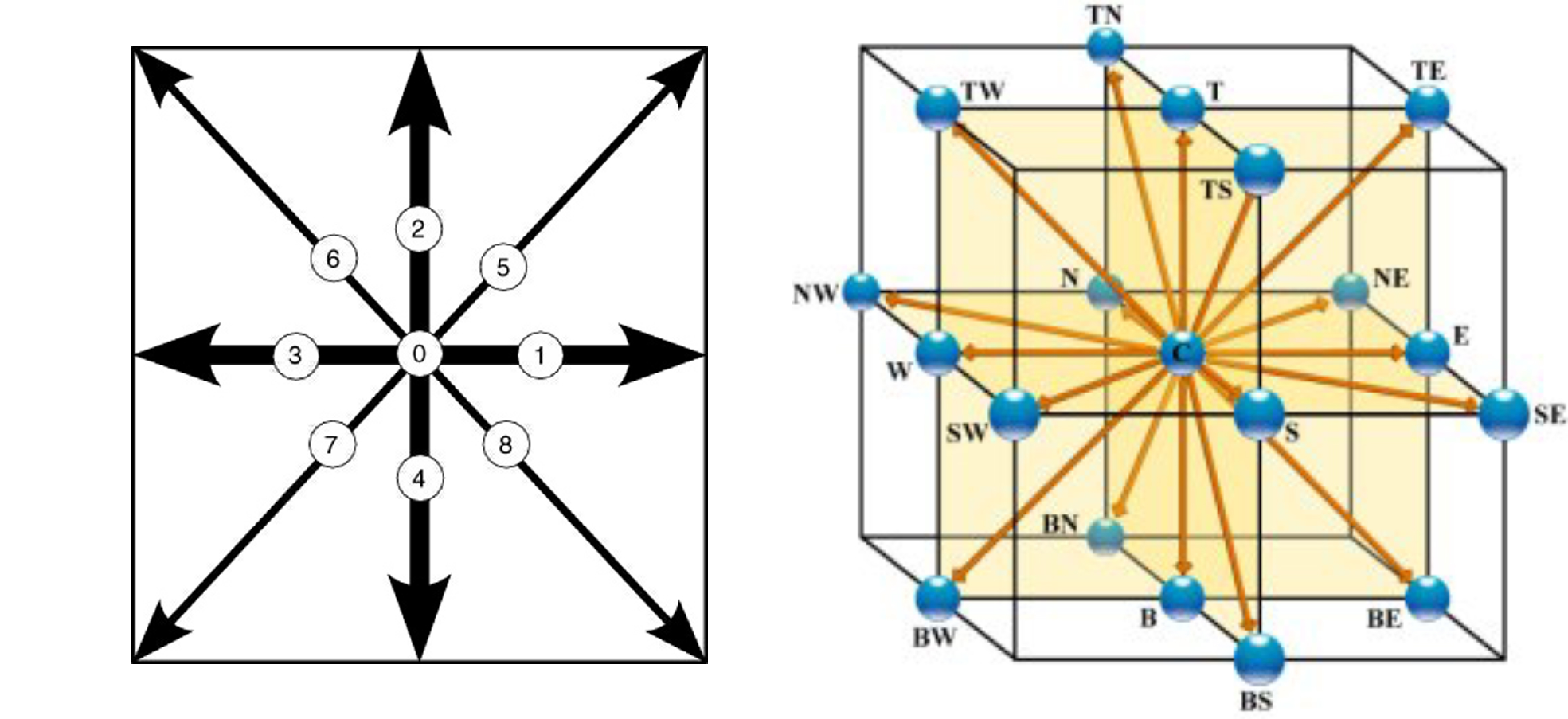}
\caption{
The D2Q9 lattice, i.e. nine discrete velocities in two spatial dimensions and the
$D3Q19$ lattice, i.e. nineteen discrete velocities
in three spatial dimensions.
These lattices provide fourth order isotropy and permit to recover the
isothermal, quasi-incompressible Navier-Stokes equations.}
\end{figure}

\subsection{Interacting fluids}

One of the main merits of the LB formalism is its flexibility towards the inclusion
of complex physics beyond the realm of Navier-Stokes hydrodynamics, through
the source term $S_i$. The typical case are force fields, representing either
coupling with the external environment, as well as internal forces resulting from
inter-molecular interactions.
A very general and fruitful expression of LB force fields is \cite{SC}:
\begin{equation}
\vec{F}(\vec{r};t) = \psi(\vec{r};t) \sum_i \vec{r}_i G_i \psi(\vec{r}+\vec{c}_i \Delta t;t) 
\end{equation}
where the sum extends to a suitable neighborhood of the lattice site $\vec{r}$ and $G_i$ is the
coupling strength of the interactions along the $i$-th link \cite{FALCU}.
A suitable tuning of the local functional $\psi(\vec{r};t) \equiv \psi[\rho(\vec{r};t)]$ permits to 
recover the main features of interacting fluids, namely non-ideal equation of state, surface tension and 
various type of dispersion forces. 

The key observation is that such forces can be reabsorbed into a 
shifted-local equilibrium of the form:
$$
f^{eq}= f_M({\vec{c}),\;\;\; \vec{c} \equiv \frac{\vec{v}-\vec{u}-\frac{\vec{F}\tau}{m}}{v_{th}}},
$$
where $f_M$ denotes the Maxwell-Boltzmann distribution and $v_{th} = \sqrt {\frac{k_BT}{m}}$ is the thermal speed.

As a result, the powerful stream-collide paradigm of the force-free formulation 
is kept intact: the new physics is simply encoded in the force shift.
This permits to incorporate a broad class of (weakly) interactions beyond hydrodynamics, without
loosing the major computational perks of the force-free formalism. 

Of course, this is limited to weakly coupled fluids, with Froude numbers well below unity,
$Fr <<1$, a situation which covers nonetheless a broad variety of soft matter flows.

In the following, we shall provide a few concrete examples drawn by recent and past
experience in the field.

\section{Chimera in action: nano and micro-flows}

LB can be taken to the nanoscale, provided fluctuations are properly reinserted
through the source term $S_i$, this time a stochastic source compliant with
the fluctuation dissipation theorem.
On the other hand, for flow quantities resulting from time and ensemble 
averaging over sufficiently long intervals as compared with the collisional scales, one may 
stay with deterministic forces expressing just the proper time-averaged interactions. 
This permits to simulate a host of interesting nano-fluidic phenomena which are
not directly amenable to a continuum description and yet too demanding for molecular dynamics.

\subsection{Super-hydrophobicity}

Super-hydrophobicity is the counterintuitive phenomenon by which fluids may experience
less friction by flowing on corrugated surfaces than on smooth ones.
The reason for this "paradox" is that, under proper physico-chemical conditions, the development of
a vapour film between the solid wall and the flowing liquid proves able of lowering 
the free-energy budget required to sustain the  three phase (Solid-Liquid-Vapour).
Formally:
$$
\sigma_{VS} S_{VS} + \sigma_{LV} S_{LV} < \sigma_{LS} S_{LS}
$$
where $\sigma_{AB}$ denotes the surface tension between phases $A$ and $B$, inclusive
of the associated contact angle and $S_{AB}$ is the area of the corresponding contact surface. 
Super-hydrophobicity was pointed out by molecular dynamics simulations of microscale 
flows over nanometric corrugations \cite{BARRAT} (see Fig. 2, left panel).
In particular a non-monotonic relation between the pressure deficit between the liquid and 
vapour phases as a function of the size of the corrugation was computed and  
shown to exhibit a loop structure, supporting a so-called de-wetting transition: the 
fluid in contact with the wall transits to the vapour phase, while the bulk remains liquid.
If the physics of super-hydrophobicity is controlled only by transport parameters such 
as surface tension and contact angle, there are good reasons to believe that molecular
dynamics should not be needed to unravel it.
On the other hand, it is not immediately obvious that the above nanoscale features can
be readily included within a continuum description.
Under such state of affairs, it may argued that a mesoscale approach, equipped
with suitable interfacial interactions, should be able to capture superhydrophobic
behaviour at a tiny fraction of the cost as compared to MD.

This is indeed the case, as first shown in  LB simulations \cite{SBRA2006}, which were able
of reproducing the pressure-geometry "equation of state" to quantitative accuracy as compared to MD data. 

\subsubsection{What did we gain?}

It is of interest to point out that such simulations required a sub-namometric 
mesh-spacing, $\Delta x_{LB} = 0.3$ nm, i.e. basically the range of molecular interaction.

One might thus wonder what have we gained in the process of using LB instead of MD.

The advantage is arguably three-fold.

First, even though the LB mesh-spacing is the same as the molecular range, the LB time-step
still is at least an order of magnitude larger:
$\Delta t_{LB} \sim \frac{\Delta x}{c_s} \sim \frac{r_0}{c_s}$
versus
$\Delta t_{MD} \sim 0.1 \frac{r_0}{c_s}$.
This is because LB is stable up to $\Delta t_{LB} <  2 \tau_c$, where $\tau_c \sim r_0/c_s$,
with mass and momentum conserved to machine accuracy, while MD requires 
a time-step much smaller than $\tau_c$ in order to secure accurate energy conservation.

Second, updating a single LB degree of freedom (the discrete population $f_i$)
is significantly faster than the update of a single MD molecule.
This is because the LB interactions are lattice-bound, hence do not require the
construction of a dynamic link-list to identify the interacting neighbours.
This counts easily another order of magnitude in favour of LB.

Finally, LB requires no statistical average, since, by definition, the distribution
function represents a large number of molecules.

Taken all together, this gives a two or three orders of magnitude LB/MD speed-up, for
the same spatial size of the problem.
In other words, LB performs a substantial coarse-graining in time.

Of course, this is {\bf not} MD, since MD would account for molecular fluctuations
completely neglected in the present LB formulation.
However, for the purpose of investigating the physics of superhydrophobic transport, such
fluctuations are inessential, thus setting a needless tax on the MD simulations.  

\begin{figure}
\centering\includegraphics[width=36pc]{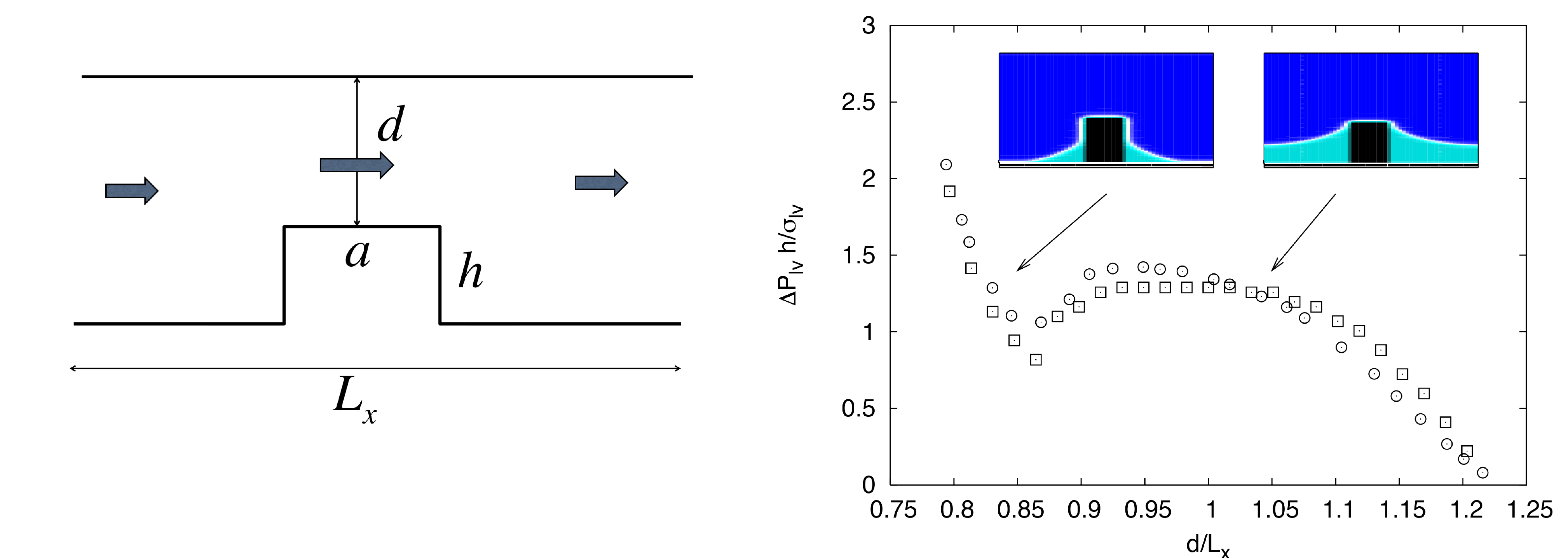}
\caption{
Super-hydrophobic nanoflow.
Left: the geometrical set-up of the flow with nano-corrugations.
Right: Pressure drop between the liquid and valor phase as a function of the
distance $d$ between the top of the nano-corrugation and the upper wall
of the micro-channel. The liquid within the nano-corrugation de-wets from 
solid wall leading to a vapour phase which sustains the liquid flowing on the top.
The liquid then flows on the vapour with less friction than on the smooth solid wall.
From \cite{SBRA2006}.
}
\label{SUPER}
\end{figure}

\subsection{Suppression of fast water transport in graphene oxide}

The same effect has been highlighted again for the study of fast water transport in graphene oxide nano flows.
While pure graphene sustains slip flow, with mass flow rates far exceeding the hydrodynamic value,
graphene oxide, exhibiting hydroxyl groups protruding from the walls, is believed to suppress such fast
water transport (FWT) regime.The debate on whether the nano flow is hydro on molecular is pursued by 
either Navier-Stokes with slip boundary condition or molecular dynamics.
In order to mimic the friction effect due to the protruding molecules,
a mesoscale LB model including a frictional interaction at the interface 
has been developed.
The frictional force takes the following form
$$
F_x(y) = \gamma_0 e^{-y/w} \;u_x(y)
$$
where $\gamma_0$ is a typical collision frequency between the water molecule
and the oxydril groups, $w$ the size of the hydroxyl molecules and $u_x$ the 
water flow along the mainstream direction $x$, $y$ being the crossflow direction.

This model proves capable of predicting the breakdown of the FWT regime
and also highlighted a subtle coexistence of collective motion in the bulk
and individual molecular motion in the near-wall region.
The LB model runs about two orders of magnitude faster than MD simulation.
\begin{figure}
\centering\includegraphics[scale=0.5]{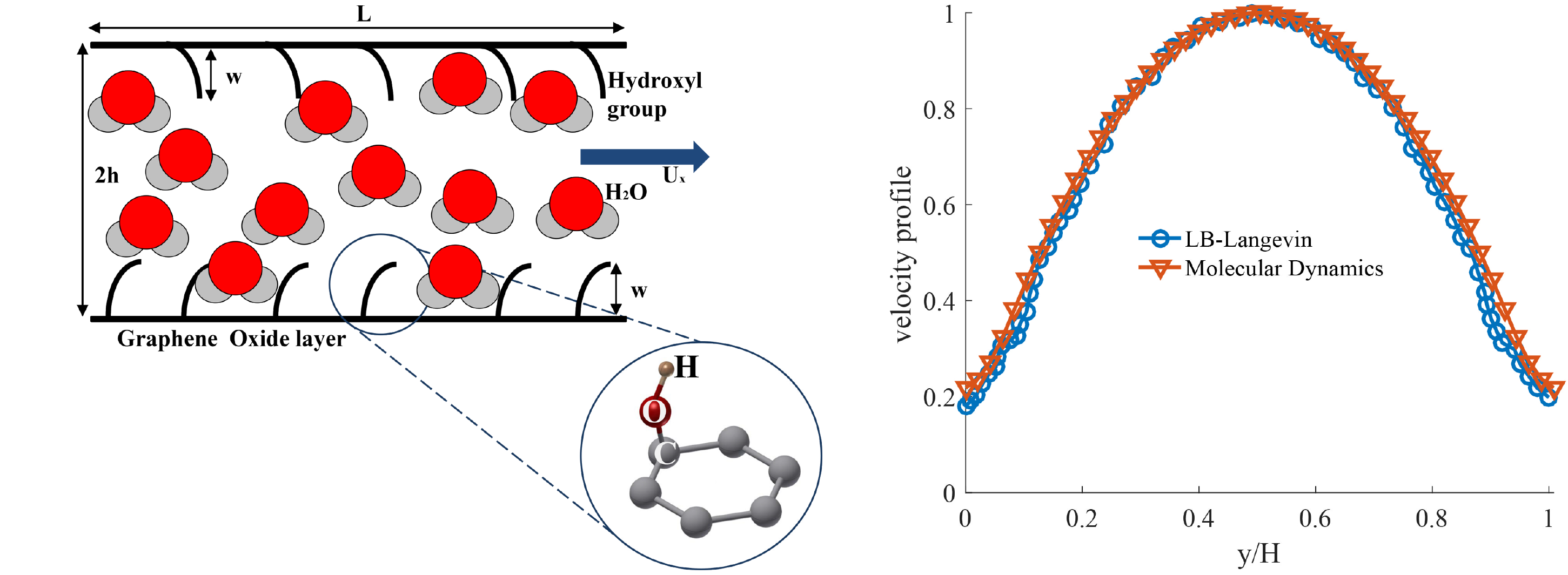}
\caption{
Examples of WBU nanoflow.
Left: Hydroxyl groups protruding from graphing oxide walls in a nanometric channel.
Right, the velocity profile computed with LB (blue) and MD (Right).
}
\label{GOX}
\end{figure}

\subsection{The Knudsen paradox}

A typical manifestation of non-equilibrium effects beyond the
hydrodynamic picture is the so called Knudsen paradox.
According to hydrodynamics, the mass flow rate should
scale inversely with the Knudsen number
$$
Q_{hydro} = \frac{1}{12} \frac{g}{\nu h^2} \propto \frac{1}{Kn} 
$$
where $g=\nabla p/\rho$ is the pressure drive
along the channel of width $h$ and the Knudsen number is defined as
$
Kn = \frac{\lambda}{h} 
$
$\lambda$ being the mean-free path.
Since the Knudsen number is proportional to the kinematic viscosity 
small Knudsen imply low dissipation hence high mass flow, and viceversa.
The viceversa part, however, is deceiving, since the above
proportionality only holds if $Kn<<1$.

Indeed, actual experiments show that beyond a critical value, around $Kn \sim 1$,
the mass flow starts to increase with the Knudsen number, a typical 
signature of individual molecular behaviour, i.e. free-streaming.

Such kind of behaviour cannot quantitatively reproduced by the 
Navier-Stokes equations, neither by the standard LB.
However, Regularized LB with higher order lattices has proven capable
of reproducing this prototypical non-equilibrium phenomena.
In a nutshell, the regularisation consists in projecting out the higher
order non-equilibrium moments after the streaming step, so as to minimise their
effect on the lower order kinetic moments governing the transport property of the
flow \cite{REG}. 

\begin{figure}
\centering
\includegraphics[scale=0.5]{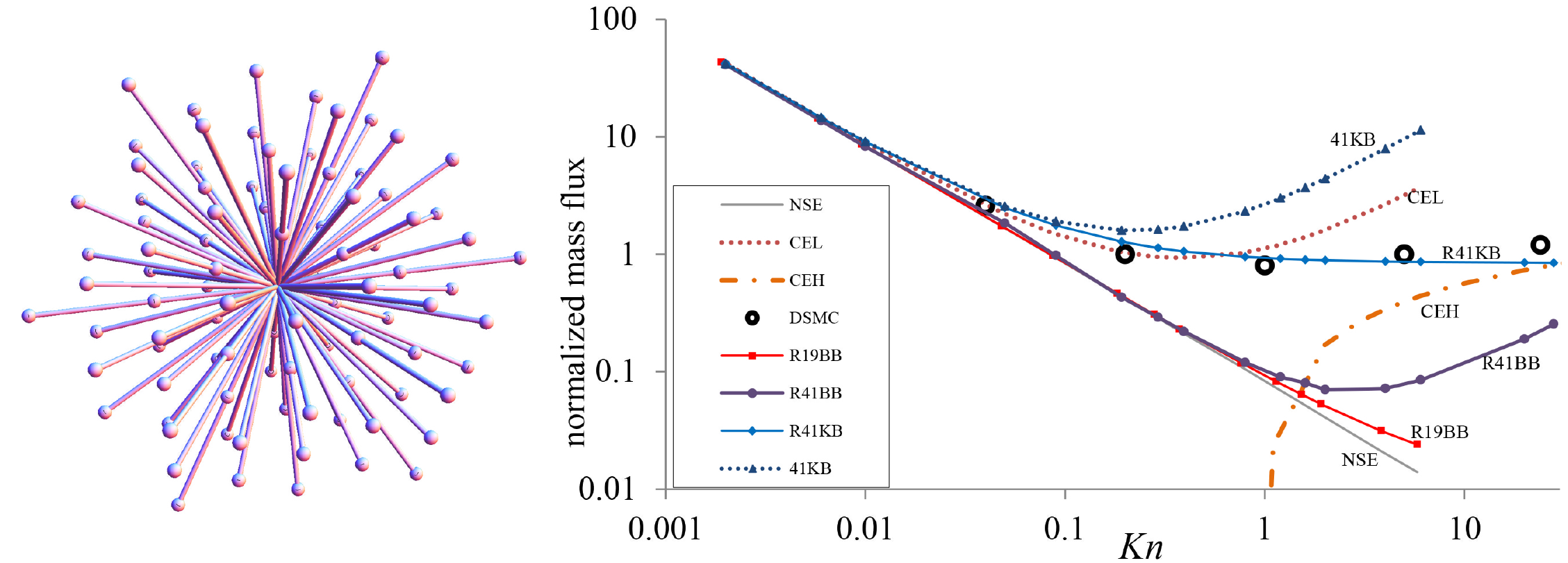}
\caption{
Left: a high order lattice with $93$ discrete speeds, featuring eighth order isotropy.
Right: the mass flow rate as a function of the Knudsen number for
a two-dimensional channel flow. The open circles refer to the 
DSMC solution, which is taken as a benchmark reference.
CEL and CEH refer to analytical asymptotic solutions obtained
by Cercignani in the low (CEL) and high (CEH) regimes, respectively.
NSE reports the Navier-Stokes solution, while the remaining curves refer
to assorted variants of the LB method: KB refers to kinetic 
boundary conditions versus bounce-back (BB).
The former allow for slip flow at the wall, while the latter does not.
The prefix R denotes the Regularized version and finally $19$ and $41$ 
refer to the number of discrete velocities.
As one can see, the R41KB model comes very close to DSMC, providing
a concrete example of LB morphing into Boltzmann.
The morphing is not perfect and it applies only to a global
quantity, the mass flow rate, but still encouraging.   
From \cite{PREMontes}.
}
\end{figure}
As one can appreciate, a suitable LB variant (R41KB, see caption) 
model comes very close to DSMC, providing a concrete example of LB morphing into Boltzmann.
The morphing is not perfect and it applies only to a global
quantity, the mass flow rate, but still encouraging.   
Further work is needed to assess whether the morphing reaches up to
the details of the spatial structure of the velocity profile within the Knudsen layer, thereby
relieving the need of the full multiscale LB/DSMC procedure described above

\section{Future perspectives}

Many fascinating challenges lie ahead of the chimera approach, all being aimed
enhancing the degree of molecular individualism within the stream-collide lattice kinetic harness.
Here we mention just a few relevant instances.

\subsubsection{Lattice BBGKY} 

The idea is to develop a lattice version of the two-body Liouville equation
for the joint distribution $f(\vec{r}_1,\vec{r}_2,\vec{v}_1,\vec{v}_2;t)$.
This would permit to replace heuristic pseudo-potentials with atomistic potentials.
The computational cost is steep: a very modest $32$ space resolution demands of the 
order of $20 \times 20 \time 32^6 \sim 0.4$ trillion degrees of freedom!
In principle this can be attempted, at least as a proof of concept.
The physical issue is how to concoct a realistic two-body local equilibrium in regions
with strong inhomogeneities, say sharp interfaces, where the two-body correlation
is neither homogeneous nor isotropic.

\subsubsection{LB with internal degrees of freedom}

Here the idea is to enrich LB with the statistical dynamics of internal
degrees of freedom, say descriptors of the molecular structure (shape) such
as angular variables.
Each internal degree of freedom adds an extra-dimension to phase-space, and
consequently economic discretizations are mandatory.

\subsubsection{LB with strong fluctuations}

Define a chimera distribution blending the Boltzmann distribution with
particles (Klimontovich distribiution),
$\Psi(\vec{r},\vec{v};t) =w \; f(\vec{r},\vec{v};t)+(1-w) \;\delta[\vec{r}-\vec{r}_p(t),\vec{v}-\vec{v}_p(t)]$.
The regular component carries the hydrodynamic content, while the singular one
takes care of non-equilibrium fluctuations, so as to 
absorb strong fluctuations which would violate the positivity of the regular one.
This is similar to standard LB/DSMC coupling, except that the weight
$w$ can be made a function of space and time, so that both regular ($f$) and singular ($\delta$)
components are evolved concurrently. 
The advantage is that the particle component an be kept much more economical
than stand-alone, since the low-order moments are in charge of the smooth part $f$.

\section{Summary}

Lattice Boltzmann is broad and flexible, it can navigate across many scales of motion with
relatively minor enrichments, all of which can be cast in the powerful stream-collide paradigm.
LB is not perfect, whenever the lattice spacing approaches relevant physical scales, close
scrutiny of lattice artefacts is necessary and higher order lattices required.
Nevertheless, it provides handy access to complex states of moving matter characterised
by the coexistence of Universality and Molecular Individualism, a hallmark of
soft matter at the Biology/Chemistry/Physics interface.

\enlargethispage{20pt}

\section*{Funding}
This work was partially  supported by the
Integrated Mesoscale Architectures for Sustainable Catalysis
(IMASC) Energy Frontier Research Center (EFRC) of the
Department of Energy, Basic Energy Sciences, Award No. DE-SC0012573.
Funding from the European Research Council under the European
Union Seventh Framework Programme (FP/2007-2013)-ERC Grant Agreement
n. 306357 (NANO-JETS) is also kindly acknowledged.

\section*{Acknowledgements}
Valuable discussions with P.V. Coveney, J.P. Boon and P. Wolynes are kindly acknowledged.
The author also wishes to thank the Solvay Foundation for supporting the Solvay Symposium
"Bridging the Gaps at the PCB interface", which inspired the present contribution.



\begin{thebibliography}{9}

\bibitem{WEINAN}  E W, 2001
Principles of multiscale modelling,
\textit{Cambridge University Press}

\bibitem{MAAD} Abraham FF, Broughton JQ, Bernstein N, Kaxiras E, 1998.
Spanning the length scales in dynamic simulation,
\textit{Computers in Physics} \textbf{12}, 538-546

\bibitem{CISE} Succi S, Filippova O, Smith G and Kaxiras E, 2001
Applying the lattice Boltzmann equation to multiscale fluid problems,
\textit{Computing in Science and Engineering},  
\textbf{3}, 26-37  

\bibitem{FILIPPA}  Filippova O, Succi S, Mazzocco F, Bella G, 2001
Multiscale lattice Boltzmann schemes with turbulence modelling,
\textit{Journal of Computational Physics},  
\textbf{170}, 812-828 

\bibitem{BGK}  Bhatnagar PL, Gross EP, Krook M, 1954.
A model for collision processes in gases 1. 
Small amplitude processes in charged and neutral one-component systems,
\textit{Physical Review}  
\textbf{94}, 511-525 

\bibitem{LBE1}  Mc Namara G, Zanetti P, 1988.
Use of the Boltzmann equation to simulate Lattice Gas Automata,
\textit{Physical Review Letters}  
\textbf{61}, 2332-35 

\bibitem{LBE2} Higuera F, Succi S, Benzi R,  1989.
Lattice gas dynamics with enhanced collisions,
\textit{EPL (Europhys. Lett.)}  
\textbf{9}, 345-349

\bibitem{LBE3} Qian YH, D'Humieres D and Lallemand P, 1992.
Lattice BGK models for Navier-Stokes equation,
\textit{EPL (Europhys. Lett.)}
\textbf{17}, 479-484

\bibitem{LB2038} Succi S, 2015.
Lattice Boltzmann 2038,
\textit{EPL (Europhys. Lett.)}
\textbf{109} 500001, 1-7

\bibitem{SC} Chen HD, and Shan X, 1993.
Lattice Boltzmann model for simulating flows with multiple phases and components,
\textit{Physical Review E}   
\textbf{47}  1815-1819  

\bibitem{FALCU} Falcucci G, Bella G, Chiatti G, Chibbaro S, Sbragaglia M and Succi S, 2007.
Lattice Boltzmann models with mid-range interactions
\textit{Communications in Computational Physics}   
\textbf{2}  1071-1084  

\bibitem{BARRAT} Cotine-Bizonne C, Barrat JL., Bouquet L,  2003. 
Low friction flows of liquid at patterned surfaces,
\textit{Nature Materials}   
\textbf{2}  1815-1819  

\bibitem{SBRA2006} Sbragaglia M., Benzi R, Biferale L, Succi S and Toschi F, 2006. 
Surface roughness-hydrophobicity coupling in micro-channel and nano-channel flows,
\textit{Physical Review Letters}, 
\textbf{97}   204503  

\bibitem{REG} Latt J, Chopard B, 2006.
Lattice Boltzmann method with regularized pre-collision distribution functions,
\textit{Mathematics and Computers in Simulation}  
\textbf{72}  165-168

\bibitem{PREMontes} Montessori A., Prestininzi P, La Rocca M and Succi S.,  
Lattice Boltzmann approach for complex non-equilibrium flows,
\textit{Physical Review E}  
\textbf{92}  043308  

\end{thebibliography}
\end{document}